\documentclass{article}
\usepackage{spconf,amsmath,graphicx}
\usepackage{multirow}

\title{ Analysing the Masked predictive coding training criterion for pre-training a Speech Representation Model
}

\name{
    Hemant Yadav\textsuperscript{\rm 1}, Sunayana Sitaram\textsuperscript{\rm 2}, Rajiv Ratn Shah\textsuperscript{\rm 1}
}
\address{
    \textsuperscript{\rm 1}IIIT Delhi, India, 
    \textsuperscript{\rm 2}Microsoft Research India, \\
    \{hemantya, rajivratn\}@iiitd.ac.in, sunayana.sitaram@microsoft.com
}

\begin{document}
%
\maketitle
\begin{abstract}

Recent developments in pre-trained speech representation utilizing self-supervised learning (SSL) have yielded exceptional results on a variety of downstream tasks. One such technique, known as masked predictive coding (MPC), has been employed by some of the most high-performing models. In this study, we investigate the impact of MPC loss on the type of information learnt at various layers in the HuBERT model, using nine probing tasks. Our findings indicate that the amount of content information learned at various layers of the HuBERT model has a positive correlation to the MPC loss. Additionally, it is also observed that any speaker-related information learned at intermediate layers of the model, is an indirect consequence of the learning process, and therefore cannot be controlled using the MPC loss. These findings may serve as inspiration for further research in the speech community, specifically in the development of new pre-training tasks or the exploration of new pre-training criterion's that directly preserves both speaker and content information at various layers of a learnt model.

\end{abstract}


\section{Introduction}
\label{sec:intro}
In recent years, there has been a lot of interest in using the self-supervised learning (SSL) technique to learn high-level representations from speech data \cite{baevski2020wav2vec2.0, hsu2021hubert, chen2022wavlm}. SSL methods work by using the input data itself to learn high-level representations that can be used to solve a variety of tasks. The training process typically involves a pre-text task, such as masked predictive coding (MPC) \cite{devlin2018bert, liu2020mockingjay}, auto-regressive predictive coding i.e., predicting future time steps based on the past \cite{brown2020languagegpt3, chung2019unsupervised}, or contrastive predictive coding (CPC) \cite{oord2018representationcpc}. The goal is to learn representations that are invariant (to irrelevant factors), disentangled, and hierarchical in nature, meaning that they can be helpful in solving different tasks simultaneously. 

Later, these SSL models can be used as a starting point to solve multiple speech tasks in two ways: (i) either by fine-tuning the model weights for each task separately, or (ii) using them as a fixed feature extractor for different downstream tasks as in the SUPERB \cite{yang2021superb} benchmark to evaluate how well the model generalizes. Later is preferred for its computation and memory efficiency. Furthermore, the authors of SUPERB benchmark show empirically that taking the weighted sum of features extracted at various layers gives better performance compared to using only the last layer feature. Top performance models in SUPERB benchmark setting are pre-trained with masked predictive coding, using the masked prediction loss (MPL), as the pre-text task. 


Given the recent success of SSL models in the speech community, it was natural for researchers to investigate what is encoded at different layers.
Two most popular approaches are: (i) learning a probing task \cite{chung2021similarity} and (ii) analyzing the representations directly \cite{pasad2021layercpc}. \cite{chung2021similarity} studies the information encoded by different pre-training techniques as a function of similarity by training several probing classifiers. On the other hand, \cite{pasad2021layercpc} studies the phonetic content in the learned representations in various layers directly.

In this paper we analyse the impact of the masked prediction loss on the encoded information. Specifically, our study focuses on investigating the information type (content or speaker) that is learned as a direct result of masked prediction loss during the pre-training step. The findings are:  
\begin{enumerate}
    \item Our experiments suggest that the MPC loss is positively correlated to the performance increase in content-based tasks, such as automatic speech recognition. In contrast, the WavLM \cite{chen2022wavlm} paper simply makes an observation on the information type encoded at each layer and does not comment on the cause of it.

    \item While the performance of speaker-based tasks decreases in relation to the increase in the performance of content-based tasks (orthogonal nature of information type). This is an indirect consequence of the learning process and can not controlled using the MPC loss. The authors of WavLM add a data augmentation to make the task harder such that it results in lower layers encoding speaker-related information (task is harder because of speaker separation). Again no correlation with the MPC loss. But an indirect result of the task being harder. A similar observation to ours. 
\end{enumerate}




\begin{figure*}[ht]
\centering
\scalebox{0.83}{
    \begin{tabular}{ccc}
      \scalebox{0.50}[0.40]{\includegraphics{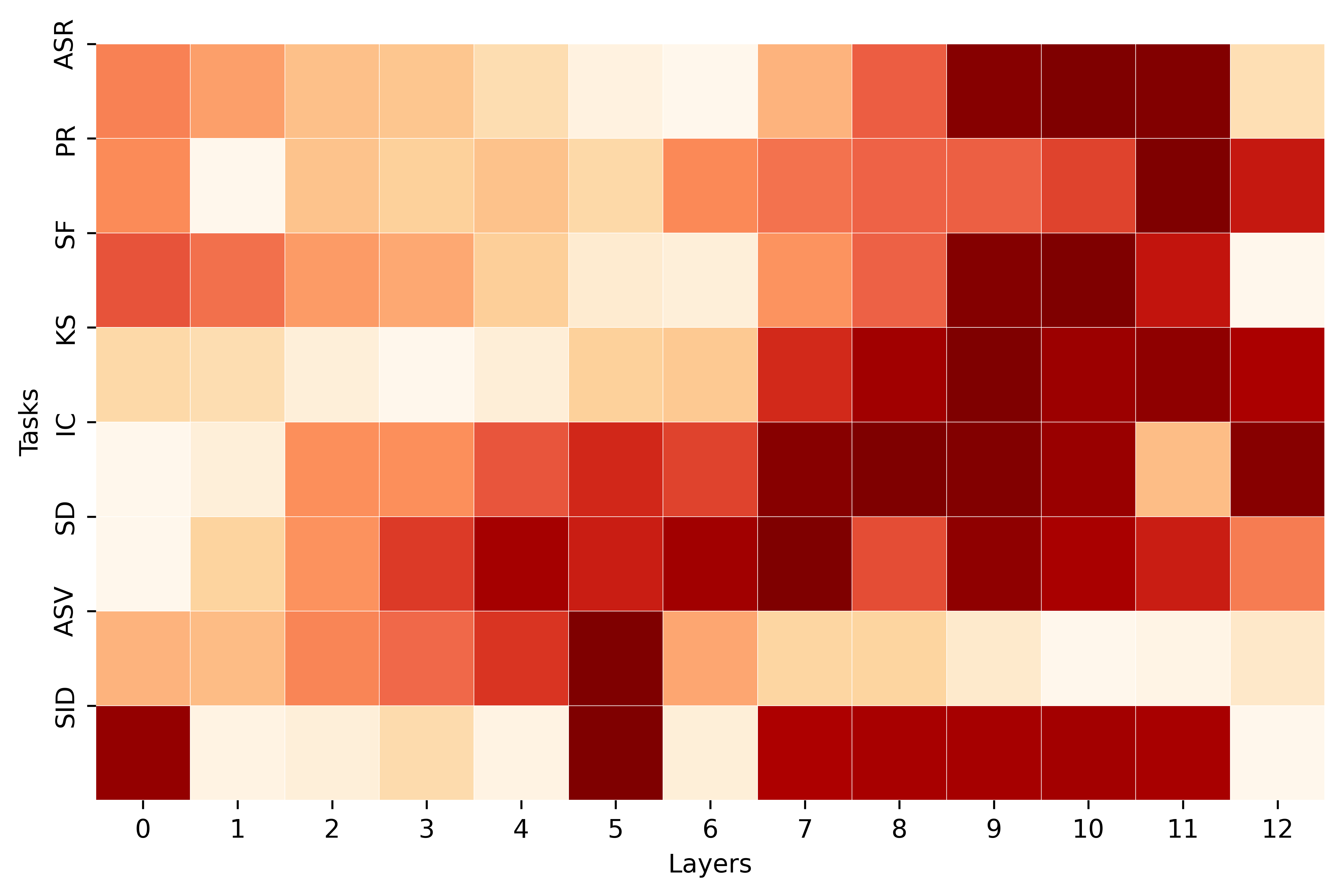}} & 
      \scalebox{0.50}[0.40]{\includegraphics{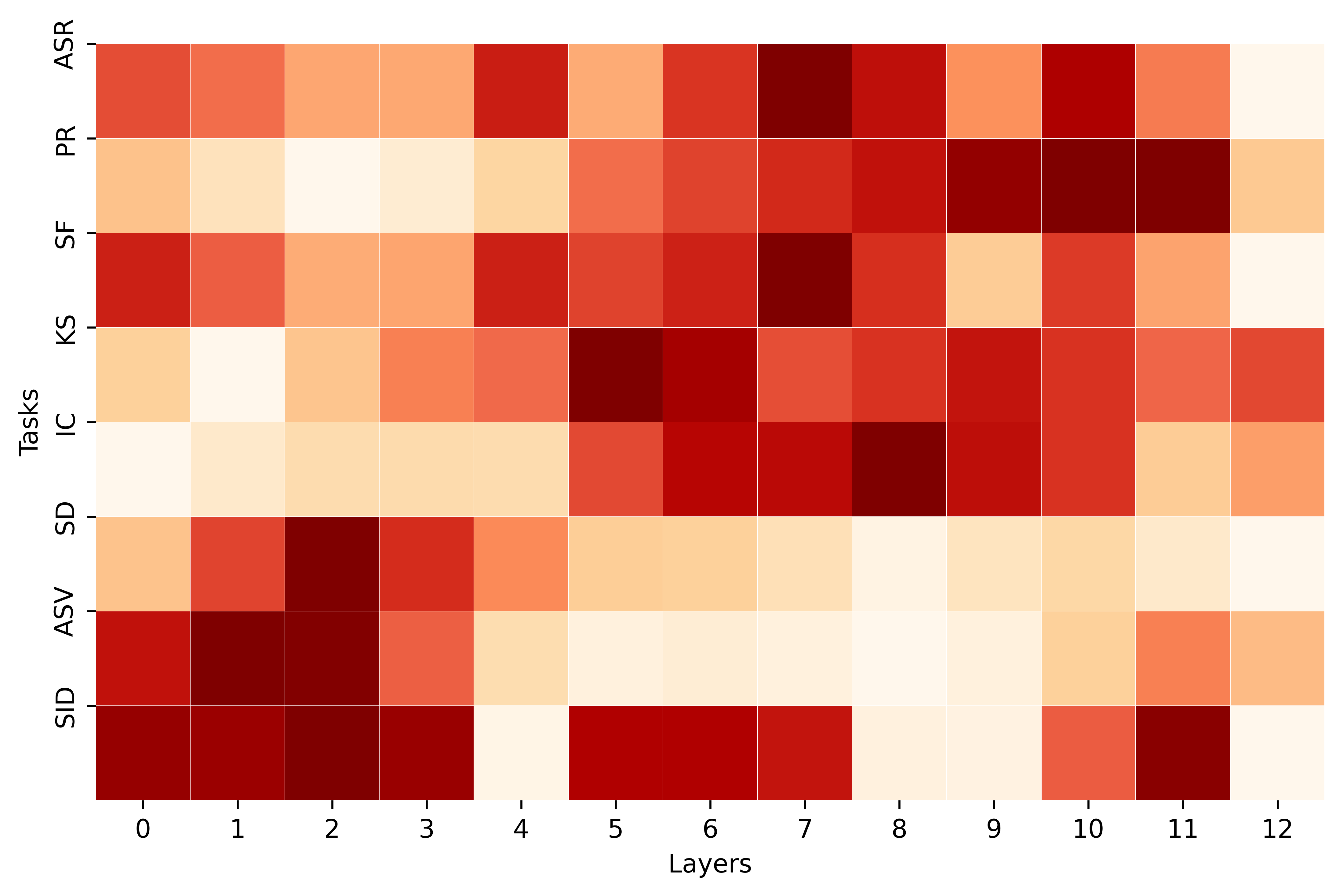}} & \scalebox{0.5}{
      \multirow{4}{*}{{
      \includegraphics{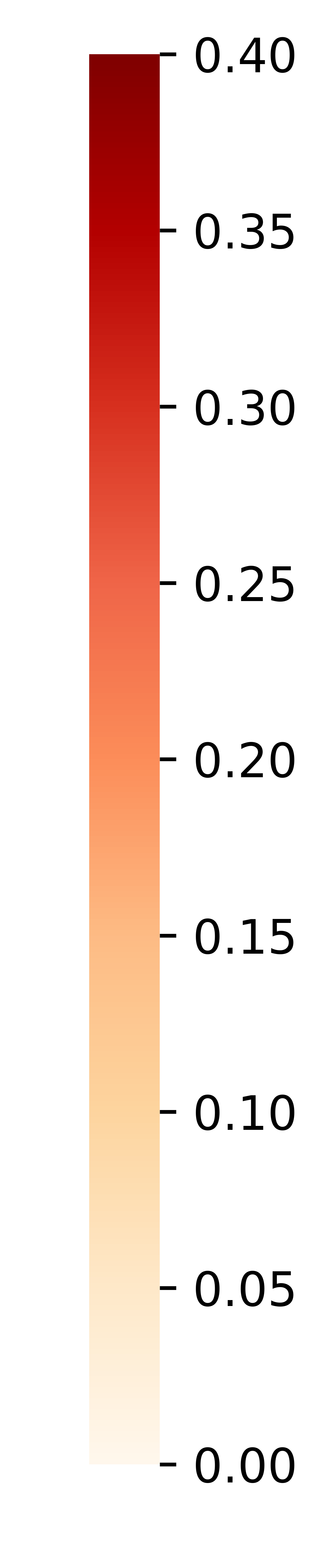}}}}
      \\
      (a) HuBERT\_1 & (b) HuBERT\_3\_0.4 & 
      \\
        \scalebox{0.50}[0.40]{\includegraphics{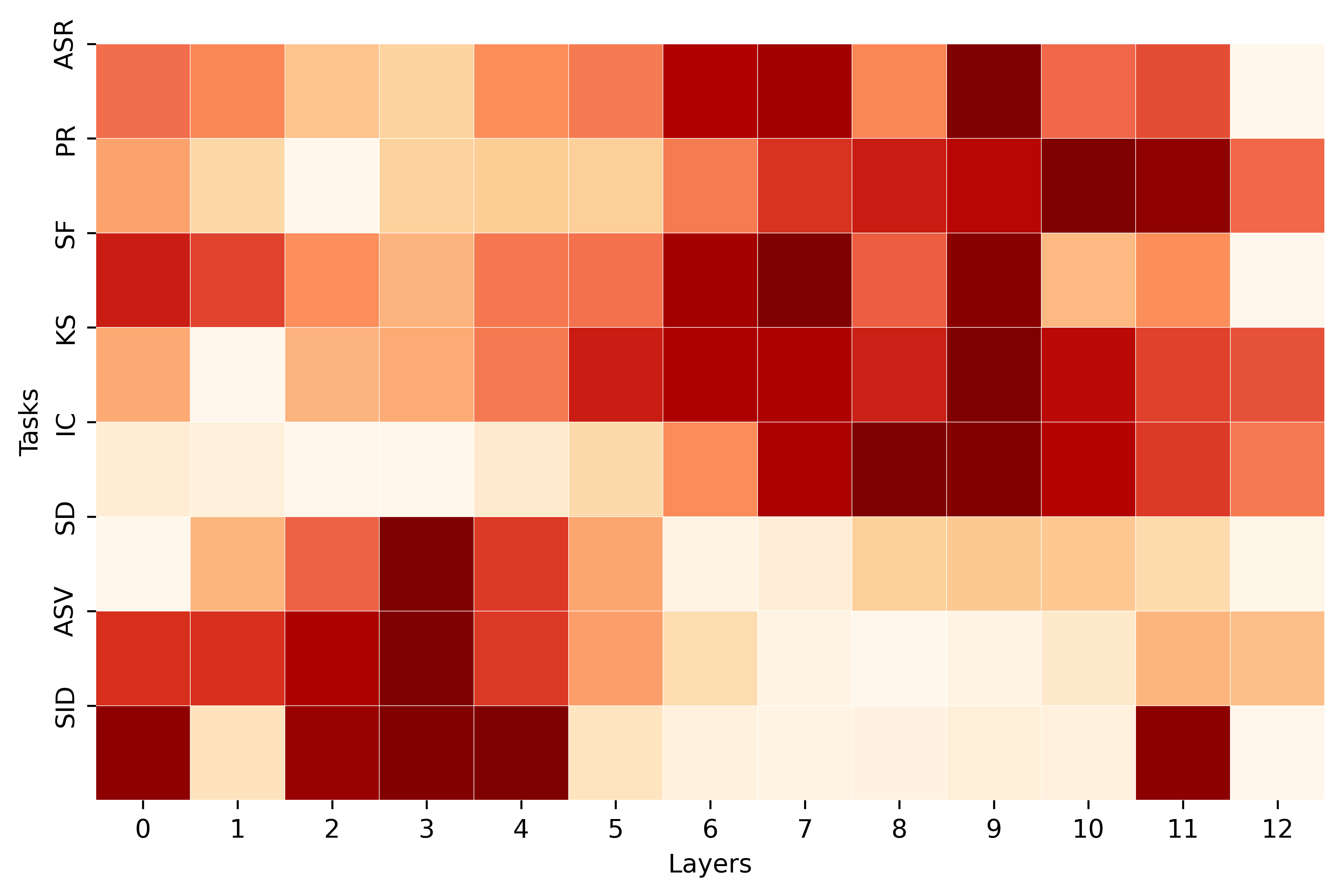}} & \scalebox{0.50}[0.40]{\includegraphics{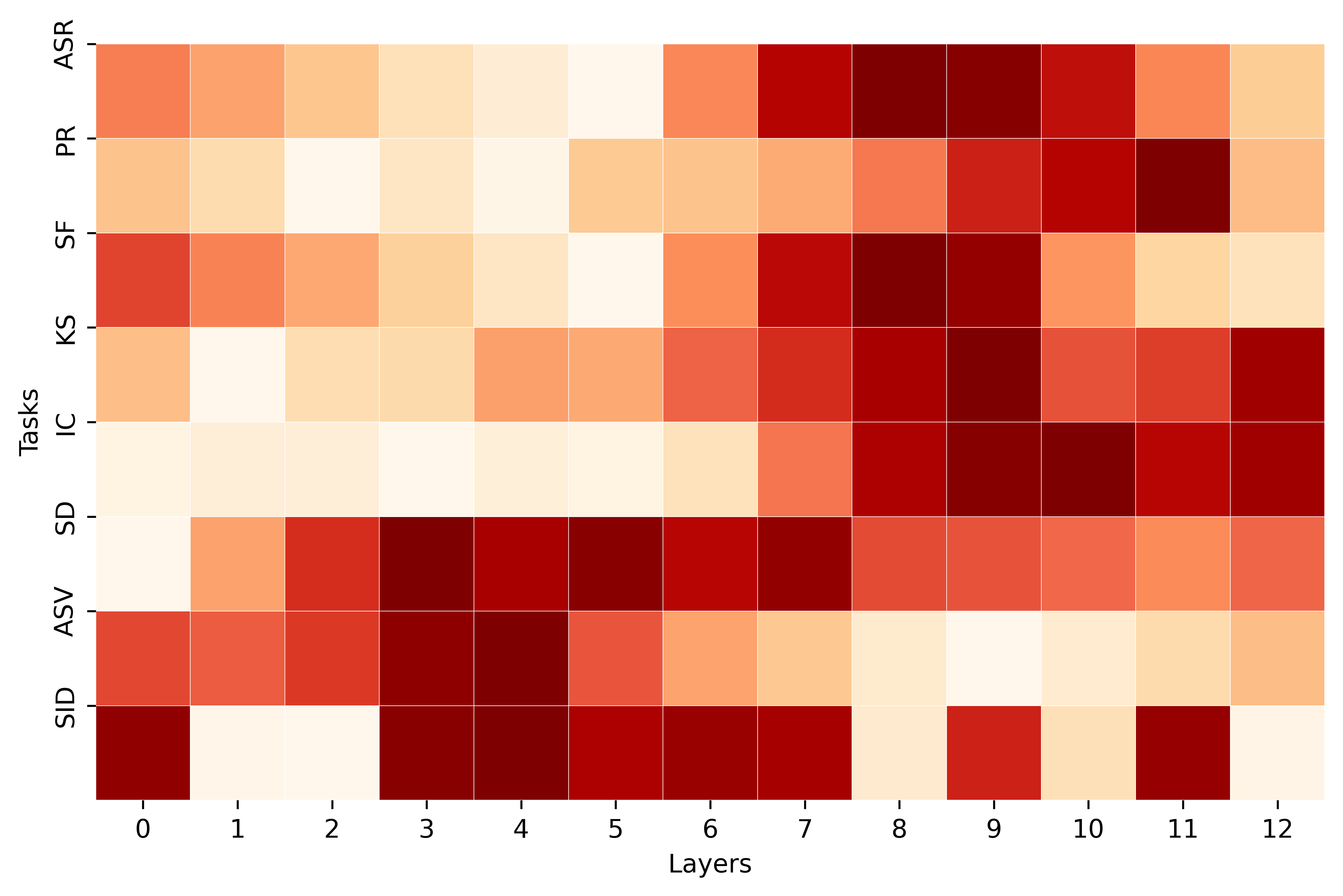}} &  \\
        (c) HuBERT\_3\_0.6 & (d) HuBERT\_3\_0.8 & \\
        \end{tabular} 
        }
\caption{Weight analysis on eight different downstream tasks, a subset of SUPERB benchmark. X-axis represents different layer numbers. Layer 0 corresponds to the input of the first HuBERT encoder layer. Y-axis corresponds eight different tasks used for evaluation. To understand the naming convention, read through Section \ref{sec:method-approach}. Better viewed in color.}
\label{figure:weightanalysis}
\end{figure*}

\section{Related Works}

As SSL models are becoming mainstream, many researchers have tried to understand what is happening under the hood of these pre-trained models either by learning a probing tasks \cite{chung2021similarity, kawakami2020learning, peng2022attention, ma2021probingacoustic} or analyzing the representations directly \cite{pasad2021layercpc} are two most popular approaches. 
The authors of \cite{pasad2021layercpc} study the WAV2VEC2.0 model \cite{baevski2020wav2vec2.0} representations  directly rather than training additional classifiers as probes. They show that  the pre-trained model follows an auto-encoder style behavior i.e., intermediate layers provide richer information of higher-level classes (phone/word information) than the initial and last layers. This is in line with other studies in the image community \cite{chen2020simplesimclr}. 

Chung et al. \cite{chung2021similarity} studies the similarity between the representations of three SSL pre-training techniques i.e., contrastive predictive coding (CPC), auto-regressive predictive coding (APC), and masked predictive coding (MPC). Their findings suggest that it is the learning objective which controls the representation similarity than architectural choices such as building blocks and directionality. Similarly, in a comprehensive review of self-supervised speech representation learning by Mohamed et al. (2022)  \cite{mohamed2022self}, the authors posited that the choice of training criterion has a greater impact on the performance gains compared to the architecture or the directionality of input.




Given pre-training criterion has the highest impact on the model's performance, in this paper we investigate the impact of the masked prediction loss on the encoded information type in various layers of HuBERT model \cite{hsu2021hubert}. To answer this we use nine probing tasks from the SUPERB benchmark. Specifically, the paper aims to determine if the use of the MPL criterion favors certain category of downstream tasks over others i.e., if the loss enforces the learned representation to be rich in content or semantic or speaker information or a combination of them.




\section{Method}
\label{sec:method}
\subsection{Manipulating the impact of the masked prediction loss training criterion on the encoded information at various layers of HuBERT model.}
\label{sec:method-approach}

HuBERT is one of the top-performing models on the SUPERB benchmark for various downstream speech tasks and is pre-trained using the MPL criterion. In brief, the pre-training process can be separated into two steps: (i) a set of pre-defined number of labels are generated using the k-means algorithm, and (ii) continuous spans of input are masked and fed to the encoder and the loss is calculated over the masked indices only. A more detailed explanation can be found in the original paper \cite{hsu2021hubert}. Given its exceptional performance on various downstream tasks, we investigate the impact of the masked prediction loss training criterion on the encoded information type at different layers of HuBERT in this study. In order to answer this question, we setup two different, yet similar, pre-training configurations using the masked prediction loss. Both the setups differ in number of loss calculations and the location where the loss is applied.

\vspace{1em}
\begin{center}
    $Total\ loss = \sum_{LOC}^{} MPL$.     
\end{center}

Where MPL is calculated similarly to \cite{hsu2021hubert} and $LOC$ is the location of it. For example, if there are 10 layers, then $LOC=\{0.4,0.6\}$ means we select the 4th and 6th layer of HuBERT to apply the MPL and sum them to calculate the total loss. 

In the first setup, similarly to \cite{hsu2021hubert}, we pre-train HuBERT by applying mask prediction loss on the final layer using 100 labels i.e., $LOC$ is 1. We refer to this setup HuBERT\_1 and is  largely similar to the original HuBERT paper, with the exception that they used a cluster size of 500. In our second setup, the number and location of MPL changes such that we apply the same loss at three different layer locations using three separate sets of labels (100, 250, 500). Here the $LOC$ has three values and therefore we sum three losses at different locations to get the final sum. 
In our second setup, the loss with 100 labels is applied to the final layer of HuBERT, with 500 labels is applied closer to the initial layers, and with 250 labels is applied equidistant from the 100 and 500 loss positions. We call this setup HuBERT\_3\_\{$LOC^{500}$\}. Where $LOC^{500}$ refers to the location of loss with 500 labels. The effect of choice of layer on which MPL with 500 labels is applied is discussed in Section \ref{sec:results}.

These two pre-training setup simulates the degree of influence of MPL on the type of information encoded at various layers. 

\subsection{Evaluating the encoded information at different layers using probing tasks.}

After the pre-training, features are extracted from each layer for evaluating their performance on nine different downstream tasks on the SUPERB benchmark and follow the standard SUPERB procedure for evaluating the two pre-training (upstream model) setups. Briefly, a linearly weighted sum of these features is used as an input to various downstream tasks, meaning different layers are more crucial for different downstream tasks. We also use these weights to draw a comparison on the information type at different layers for the two pre-training setups outlined in Section \ref{sec:method-approach} i.e., content or speaker related.


\section{Experimental details}
\noindent \textbf{Dataset}:For pre-training HuBERT (the upstream SSL model), we use the standard Librispeech 960 hours dataset. 

Nine downstream tasks are chosen from the SUPERB benchmark. These nine downstream tasks are divided into three categories (i) 3 speaker tasks:- Speaker identification (SID), Automatic speaker verification (ASV), and speaker diarization (SD) (ii) 4 content tasks:- Phoneme recognition (PR), Automatic speech recognition (ASR), Keyword spotting (KS), and Query-by-Example (QbE) (iii) 2 Semantic tasks:- Intent classification (IC) and slot filling (SF). For details, refer \cite{yang2021superb}.

\noindent \textbf{Pre-training}: As explained in Section \ref{sec:method-approach}, two models are pre-trained using MPL training criterion. In both setups, we pre-train the BASE HuBERT model on 960 hours of Librispeech data using 16 GPUs in total with a batch size of at most 78.12 seconds of audio per GPU for 400k iterations. It took around 18 hours to complete 100 thousand steps for the first setup and 24 hours for the second setup. For reference, in the original HuBERT paper, the authors trained on 32 GPUs with a batch size of at most 87.5 seconds of audio per GPU. 
Pre-trained models and training configurations can be found on GitHub \footnote{https://github.com/raotnameh/hubert\_cluster}.

\noindent \textbf{Downstream tasks}: The two pre-training setup are evaluated on nine different downstream tasks taken from SUPERB benchmark. Adhering to the SUPERB benchmark evaluation, features from each layer are linearly weighted and is used as an input for different downstream task. These weights for each task are then plotted as a heatmap, as shown in Figure \ref{figure:weightanalysis}. We use the evaluation scripts provided in the official GitHub\footnote{https://github.com/s3prl/s3prl} repository of SUPERB benchmark.

\begin{table*}[ht]

\caption{Performance of the two pre-training setups on nine different downstream tasks, a subset of SUPERB benchmark. To understand the methods naming convention, read through Section \ref{sec:method-approach}. For the SID task, unlike the SUPERB benchmark, we used a learning rate scheduler.}
\centering
    \scalebox{0.95}{
    \begin{tabular}{|p{3cm}|p{1cm}|p{1cm}|p{1cm}|p{1cm}|p{1.1cm}|p{1cm}|p{1.5cm}|p{1cm}|p{1cm}|p{1.1cm}|}
    \hline
      \multicolumn{1}{|c|}{\multirow{3}{*}{\textbf{Methods}}}  &  \multicolumn{3}{|c|}{\textbf{ Speaker}} & \multicolumn{4}{|c|}{\textbf{Content}} & \multicolumn{3}{|c|}{\textbf{Semantics}} \\
      \cline{2-11}
       & ASV & SID & SD & PR & ASR & KS & QbE & IC & \multicolumn{2}{|c|}{SF} \\
       \cline{2-11}
        & EER $\downarrow$ & Acc $\uparrow$ & DER $\downarrow$ & PER $\downarrow$ & WER $\downarrow$ & Acc $\uparrow$ & MTWV $\uparrow$ & Acc $\uparrow$ & F1 $\uparrow$ & CER $\downarrow$ \\
        \cline{1-11}
         HuBERT\_1 & \textbf{5.25} & \textbf{84.58} & \textbf{5.73} & 5.07 & 6.82 & 96.26 & \textbf{0.0999} & 96.25 & 88.7 & 25.4 \\
         HuBERT\_3\_0.4 & 5.56 & 78.80 & 6.78 & \textbf{4.25} & \textbf{5.98} & 95.60 & 0.0830 & \textbf{98.28} & \textbf{88.7} & \textbf{23.61} \\
         HuBERT\_3\_0.6 & 5.32 & 82.74 & 5.86 & 4.26 & 6.21 & \textbf{96.43} & 0.0860 & 98.10 & 88.7 & 24.95 \\
        HuBERT\_3\_0.8 & 5.18 & 84.28 & 6.03 & 4.35 & 6.80 & 96.10 & 0.0910 & 98.5 & 88.7 & 24.24 \\
        
        \cline{1-11}
        HuBERT\_3\_0.4\_stable & 5.69 & 78.71 & 6.53 & \textbf{4.16} & \textbf{5.67} & 95.94 & 0.0973 & 98.25 & 88.3 & 24.29 \\
        \cline{1-11}
        HuBERT\_1\_18 & 6.52 & 81.48 & 4.82 & 4.23 & 6.29 & 96.91 & 0.0978 & 98.5 & 89.01 & 23.7 \\
        HuBERT\_3\_0.4\_18 & 4.04 & 83.1 & 6.06 & 3.89 & 5.21 & 96.49 & 0.0670 & 98.76 & 89.16 & 22.86 \\
        \hline
        
        \end{tabular}}
\label{table:superb}
\end{table*}

\section{Results}
\label{sec:results}

\subsection{Impact of the MPC loss on the type of information learnt at various layers.}

We analyse the linearly weighted sum of layers for different downstream tasks. This gives us a better understanding of the correlation between the information type at each layer and the masked prediction loss pre-training criterion. The weights show the importance of features extracted from each layer for eight downstream tasks, as shown in Figure \ref{figure:weightanalysis}.

The first observation is that the aggressive application of masked prediction loss in the HuBERT\_3\_0.4 setup, increases the participation of each layer in the weighted sum for the content based task such as ASR and PR. In other words, a lot more layers are now active and the weights are more evenly distributed compared to HuBERT\_1. This could be a contributing factor for the improvement in performance of ASR and PR. A similar trend is also observed in the semantic-based tasks such as SF and IC. The second observation is that aggressively applying the masked prediction loss, in HuBERT\_3\_0.4, pushes the participation of layer for speaker-related task towards the few initial layers (up to layer 3). Compared to HuBERT\_1 where most of the intermediate layers are active for the ASV and SID task.

Combining these two observations, we hypothesize that the amount of content-related information learned at various layers of the HuBERT model is directly proportional to the minimization of the MPL i.e., positive correlation. An increase in the performance of content related tasks would result in a performance drop for the speaker related tasks (orthogonal nature of information type). We observe the same in our experimental analysis, as shown in Table \ref{table:superb}.

To further test our hypothesis, we pre-train two additional models, following the second setup, such that the location of masked prediction loss is now closer to the final layer with location values of 0.6 and 0.8. We observe a similar trend that the content information shifts towards the final layers, which is expected. Which results in performance degradation for ASR and PR tasks and improvements for the speaker-based tasks such as SID and ASV. Our experimental results validate the same, as shown in Table \ref{table:superb}. These two experiments further supports our initial hypothesis that the masked prediction loss and the content information in the learned representations has a positive correlation. 

\subsection{Quantitative analysis of HuBERT\_1 and HuBERT\_3\_0.4 performance on SUPERB benchmark}

We evaluate the two pre-training setups across nine different tasks, as shown in Table \ref{table:superb}. Firstly, we observe that aggressively applying the masked prediction loss at three different layers gives a boost in performance to content-based tasks and a drop in performance for speaker-based tasks and vice-versa. We also observe a similar trend for semantics-based tasks, i.e, HuBERT\_3\_0.4 either improves the performance or matches HuBERT\_1. Based on our experiments of the QbE task, we observe the decrease in performance as we increase the number of labels. In the original HuBERT paper, the authors used 500 labels and achieve a score of $ 0.0736$. Our results show that the performance of QbE task is inversely proportional to the number of labels.


\subsection{Discussion}

We also studied the impact of the cluster assignment strategy on the performance of HuBERT\_3\_0.4. Previously, to generate the labels, k-means was run randomly on a 10\% subset of data, each time to create three different clusters of 100, 250, and 500. This cluster assignment setup is referred to as CA1. In contrast, the new cluster assignment setup, referred to as CA2, involves running k-means to obtain 500 cluster points, then using these points as input to k-means again to get 250 cluster points, and repeating the process to obtain 100 cluster points. This setup is similar to the Hierarchical clustering algorithm \cite{cohen2019hierarchical}. The pre-training process explained in Section 3 is conducted again using the labels created with CA2. The results show that the performance gains for ASR and PR tasks are even greater when using labels generated from CA2 compared to CA1 as shown in Table \ref{table:superb}. It is suggested that the labels generated with CA2 setup are more stable, meaning the behavior in the middle layer embeddings learned from 500 and 250 labels support/align with the 100 labels learned in the last layer. There is a severe degradation on ASV and SID tasks, requiring informaiton orthognal to the ASR and PR tasks. The reason for this behavior is unknown and left for future research. 

Lastly, we pre-train a deeper model (18 layers) using the CA1 setup and observe that only the ASR and PR performance has increased for the HuBERT\_3\_0.4 compared to the HuBERT\_1. This again proves our hypothesis that the content information encoded in various HuBERT layers and the MPC loss has a positive correlation.

\section{Conclusion and future work}
In this work, we analyze the effect of MPL pre-training criterion on what is encoded at different layers. We find that minimizing the MPL maximizes the content information in the layers closer to where the loss is calculated. While at the same time forgetting speaker information and pushing them far from the layer where loss is calculated. This is true in general also, i.e., information require to solve the ASR and SID task is orthogonal. Based on our findings we hypothesize that simply applying MPL to learn speech representations is not the way forward for pre-training ONE universal model that can be used for different downstream tasks. 


\section{Acknowledgments}
Rajiv Ratn Shah was partly supported by the Infosys Center for Artificial Intelligence and the Center of Design and New Media at IIIT Delhi, India. Hemant Yadav is supported by Microsoft Research India PhD Fellowship program.

\bibliographystyle{IEEEbib}
\bibliography{main}

\end{document}